\documentclass[twocolumn,showpacs,preprintnumbers,amsmath,amssymb]{revtex4}

\usepackage{graphicx}
\usepackage{dcolumn}
\usepackage{bm}

\begin{document}

\preprint{}

\title{Magneto-optics of two-dimensional electron gases modified by strong Coulomb interactions in ZnSe quantum wells }

\author{D.~Keller$^{1}$}
\author{D.~R.~Yakovlev$^{2,3}$}
\author{G.~V.~Astakhov$^{1,3}$}
\author{W.~Ossau$^{1}$}
\author{S.~A.~Crooker$^{4}$}
\author{T.~Slobodskyy$^{1}$}
\author{A.~Waag$^{5}$}
\author{G.~Schmidt$^{1}$}
\author{L.~W.~Molenkamp$^{1}$}

\affiliation{ $^{1}$Physikalisches Institut der Universit\"{a}t
W\"{u}rzburg, 97074 W\"{u}rzburg, Germany \\
$^{2}$Experimentelle Physik 2, Universit\"{a}t Dortmund, 44221
Dortmund, Germany \\
$^{3}$A.F.Ioffe Physico-Technical Institute, Russian
Academy of Sciences, 194021 St.Petersburg, Russia \\
$^{4}$National High Magnetic Field Laboratory, Los Alamos, New
Mexico 87545, USA \\
$^{5}$Institute of Semiconductor Technology, Braunschweig
Technical University, 38106 Braunschweig, Germany }

\date{\today}

\begin{abstract}
The optical properties of two-dimensional electron gases in
ZnSe/(Zn,Be)Se and ZnSe/(Zn,Be,Mg)Se modulation-doped quantum
wells with electron densities up to $1.4 \times 10^{12}$~cm$^{-2}$
were studied by photoluminescence, photoluminescence excitation
and reflectivity in a temperature range between 1.6 and 70~K and
in external magnetic fields up to 48~T. In these structures, the
Fermi energy of the two-dimensional electron gas falls in the
range between the trion binding energy and the exciton binding
energy. Optical spectra in this regime are shown to be strongly
influenced by the Coulomb interaction between electrons and
photoexcited holes. In high magnetic fields, when the filling
factor of the two-dimensional electron gas becomes smaller than
two, a change from Landau-level-like spectra to exciton-like
spectra occurs. We attempt to provide a phenomenological
description of the evolution of optical spectra for quantum wells
with strong Coulomb interactions.
\end{abstract}


\maketitle


\section{\label{sec1} INTRODUCTION }

Optical spectroscopy is a powerful tool to study the energy
spectrum of electronic states in low-dimensional heterostructures.
It was successfully used for investigation of two-dimensional
electron gases (2DEG) (for review see e.g. Ref.~\onlinecite{1a}),
providing additional information not available from transport
measurements alone. However, in optical experiments the presence
of photoexcited holes perturbs the energy spectrum of the 2DEG via
electron-hole Coulomb interactions, an effect which should be
taken into account for proper interpretation of optical spectra.
This is already the case for III-V semiconductor heterostructures
based on GaAs, where the Coulomb interaction is relatively weak
(e.g. the exciton binding energy in bulk GaAs is 4.2~meV).
However, Coulomb interactions are of major importance for II-VI
semiconductors (e.g. the exciton binding energy in bulk ZnSe is
20~meV). Modification of the optical spectra of modulation-doped
quantum wells (QW) containing a 2DEG has been the subject of
extensive theoretical \cite{1,2,3,4,5,6,7,8,8a} and experimental
studies \cite{9,10,11,12,13,14,15,15a,
16,17,18,18a,18b,29,30,31,32}.

In undoped QWs, the ground state of an optically excited single
electron-hole pair is an exciton ($\mathrm{X}$), which is formed
by the Coulomb interaction between the electron and the hole. For
QWs with a very dilute two-dimensional electron gas (2DEG) where
the electron concentration $n_e  \ll \frac{1}{\pi {a_B}^2}$ (here
$a_B$ is the radius of the quasi-two-dimensional exciton), a
negatively charged exciton complex is the energetically lowest
excitation \cite{19,20,20a,20b}. The negatively charged exciton is
composed of two electrons and a hole, and is commonly called a
trion ($\mathrm{T}$), and is analogous to  the negatively charged
$\mathrm{H^-}$ state of the hydrogen atom. In semiconductor QWs
trions have binding energies of a few meV and are observed in
optical spectra below the exciton resonance. In the limit of zero
electron concentration, the energy separation between trion and
exciton optical transitions corresponds to the binding energy of
"isolated" trions  $E_B^T$ (i.e. the trion state which is not
perturbed by interaction with the 2DEG). It varies usually from 5
to 20\% of the exciton binding energy $E_B^X$ depending on the QW
thickness and the height of the barriers (see e.g.
Ref.~\onlinecite{21calc} and references therein).

For a high density 2DEG with  $n_e  \geq \frac{1}{\pi {a_B}^2}$,
screening and phase space filling effects destroy the strong
Coulomb attraction between an individual electron and hole, i.e.
the exciton is suppressed \cite{1}. As a result, instead of a
narrow excitonic emission, a broad emission band with a linewidth
determined by the 2DEG Fermi energy $\varepsilon_F$ is
characteristic of this regime. Also, a Fermi-edge singularity
($\mathrm{FES}$) appears in optical spectra owing to the Coulomb
interaction of a photocreated hole with the electrons at the
Fermi-surface \cite{4,9}. Optical spectra are usually interpreted
in terms of band to band transitions, dressed with many-body
interactions of the excited or annihilated electron-hole pair with
electrons of the Fermi sea \cite{1}. The energies of the emission
and absorption thresholds are shifted following to a band-gap
renormalization. Extended numerical calculations are needed to
model optical spectra of a QW with a dense 2DEG. On the basis of
such calculations Hawrylak \textit{et al.} predicted the presence
of two thresholds (two peaks) in the absorption spectrum
corresponding to the bound states of the photoexcited hole
interacting with a 2DEG. The first threshold occurs at an energy
$\hbar \omega_1$ and is a doubly occupied trion complex. The
second threshold with an energy $\hbar \omega_2$ corresponds to an
exciton created in the 2DEG \cite{5,6,34}. This approach allows to
explain the main features in the optical spectra of modulation
doped GaAs QWs \cite{6,7}.

Evolution of the optical spectra with increasing electron density
from the low-density to the high-density limit is an important
issue which has been examined experimentally in recent years for
GaAs [33], CdTe \cite{14,15,18a,18b} and ZnSe \cite{18,29} based
QWs. General trends have been established and attempts to
formulate a phenomenological description have been undertaken. It
has been shown that for all studied material systems the filling
factor $\nu = 2$ separates Landau-level-like behavior from an
excitonic-like behavior in external magnetic fields. In high
magnetic fields for $\nu < 2$ the broad emission band transforms
into a narrow line that shifts diamagnetically, reminiscent of the
trion line in low-doped samples. Further field increase to $\nu <
1$ revives one more line in optical spectra, which is excitonic
like. In these fields, optical spectra of moderately-doped and
very weakly-doped QWs look identical. Despite the considerable
progress in experiment the situation is still far from completely
understood, in major part due to the absence of a proper model
description of the intermediate concentration regime. In this
regime with  $\frac{0.05}{\pi {a_B}^2} \leq n_e  \leq \frac{1}{\pi
{a_B}^2}$ the Fermi energy of the 2DEG falls in the range between
the trion and exciton binding energies  $E_B^T \leq \varepsilon_F
\leq E_B^X$, and the effect of the Coulomb interaction on the
optical spectra can be discussed in terms of comparing these three
characteristic energies.

ZnSe-based QWs exhibit very strong Coulomb electron-hole
interactions (the exciton binding energy is twice as large as
compared to CdTe QWs and four times larger than in GaAs QWs). The
Coulomb energy amounts to 30-40~meV. In earlier studies of n-type
doped (Zn,Cd)Se/ZnSe QWs the well layers were made of ternary
alloy (Zn,Cd)Se \cite{30,31,32}. This results in a considerable
broadening of optical lines by alloy fluctuations which obscures
fine details in the spectroscopic analysis.

In this paper we present a comprehensive optical study of n-type
modulation-doped ZnSe/(Zn,Be,Mg)Se QWs, with an aim to investigate
the modification of optical spectra in structures with strong
Coulomb interactions, and at the intermediate electron densities
satisfying the condition  $E_B^T \leq \varepsilon_F \leq E_B^X$.
Photoluminescence (PL), PL excitation (PLE) and reflectivity
spectra have been measured in high magnetic fields up to 48~T. The
paper is organized as follows: In section~\ref{sec2} details of
the sample structure and the experimental techniques are
described. In section~\ref{sec3} optical spectra at a zero
magnetic field are discussed. The modification of optical spectra
in high magnetic fields is presented in section~\ref{sec4}.


\section{\label{sec2}EXPERIMENTAL DETAILS}

\begin{table}
\caption{\label{tab1}Parameters of the samples containing a ZnSe
QW.}
\begin{ruledtabular}
\begin{tabular}{ccccc}
sample & barrier material & barrier & electron & Fermi \\
& & gap & density & energy \\
& & ${E_g}^b$ [eV] & [$\mathrm{cm}^{-2}$] & $\varepsilon_F$ [meV] \\
\hline \#1R & Zn$_{0.82}$Be$_{0.08}$Mg$_{0.10}$Se & 3.06 & $3
\times 10^{10}$ & 0.5 \\
\#1D & Zn$_{0.82}$Be$_{0.08}$Mg$_{0.10}$Se & 3.06 & $5
\times 10^{11}$ & 7.7 \\
\#2R & Zn$_{0.94}$Be$_{0.06}$Se & 2.93 & $8
\times 10^{10}$ & 1.2 \\
\#2D & Zn$_{0.94}$Be$_{0.06}$Se & 2.93 & $1.4
\times 10^{12}$ & 21.5 \\
\end{tabular}
\end{ruledtabular}
\end{table}

ZnSe/(Zn,Be,Mg)Se quantum wells were grown at the University of
W\"{u}rzburg by molecular-beam epitaxy on (100)-oriented GaAs
substrates. Detailed growth parameters for these structures can be
found in Ref.~\onlinecite{21}. We list some of them here for
convenience and some parameters are collected in Table~\ref{tab1}.
The band gap of ZnSe at a liquid helium temperature is 2.82~eV,
the band gaps of the barrier materials ${E_g}^b$ are given in the
Table. The band gap discontinuity between the QW and barrier is
distributed between conduction and valence band in the proportion
$\Delta E_C /  \Delta E_V = 78 / 22$. The electron effective mass
in ZnSe is $m_e = 0.15 m_0$ and the heavy-hole in-plane mass is
$m_h = 0.46 m_0$. The exciton binding energy is $E_B^X \approx
30$~meV and the trion binding energy is $E_B^T \approx 5$~meV.

\begin{figure}[tbp]
\includegraphics[width=.35\textwidth]{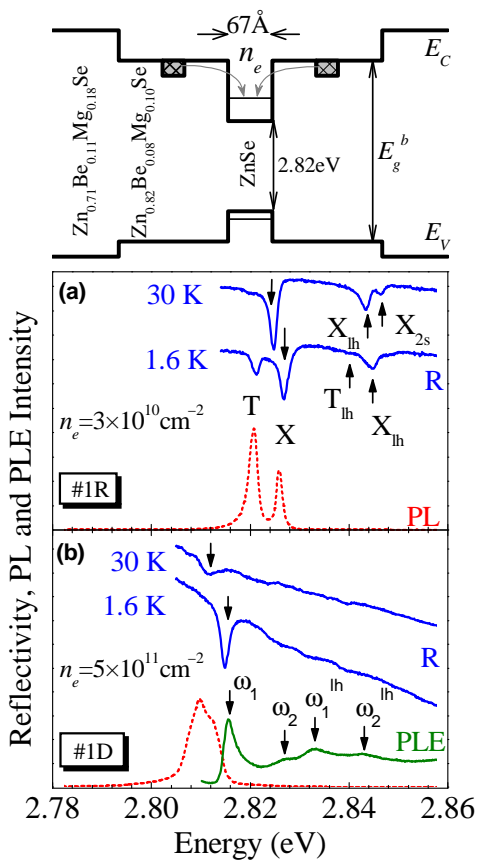}
\caption{(Color online) Photoluminescence (PL), photoluminescence
excitation (PLE) and reflectivity (R) spectra of two 67-\AA-thick
ZnSe/Zn$_{0.82}$Be$_{0.08}$Mg$_{0.10}$Se QWs, with different
electron densities: (a) reference QW \#1R with $n_e = 3 \times 10
^{10}$~cm$^{-2}$, (b) doped QW \#1D with  $n_e = 5 \times 10
^{11}$~cm$^{-2}$. $B=0$~T and $T=1.6$~K. Scheme shows a sketch of
samples type \#1.} \label{fig_I}
\end{figure}

\begin{figure}[tbp]
\includegraphics[width=.35\textwidth]{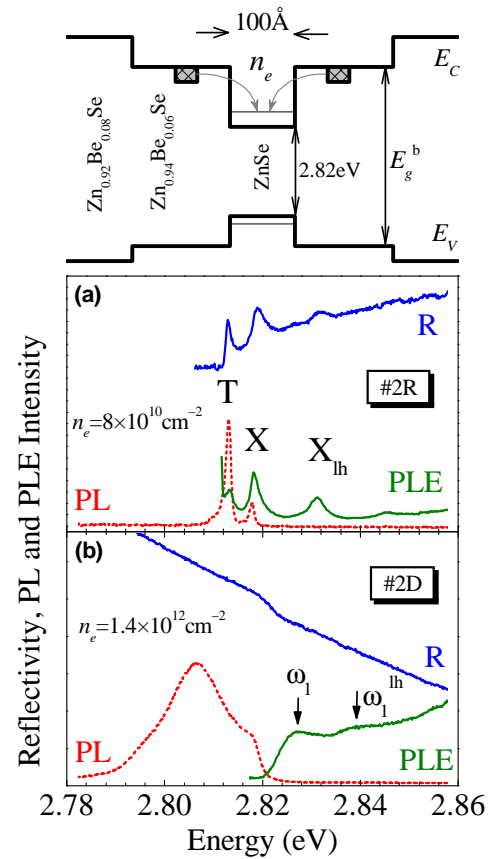}
\caption{(Color online) Photoluminescence (PL), photoluminescence
excitation (PLE) and reflectivity (R) spectra of two 100-\AA-thick
ZnSe/Zn$_{0.94}$Be$_{0.06}$Se QWs, with different electron
densities: (a) reference QW \#2R with $n_e = 8 \times 10
^{10}$~cm$^{-2}$, (b) doped QW \#2D with  $n_e = 1.4 \times 10
^{12}$~cm$^{-2}$. $B=0$~T and $T=1.6$~K. Scheme shows a sketch of
samples type \#2.} \label{fig_II}
\end{figure}

Two sets of samples have been studied, each consists of a
modulation-doped QW labeled with "D" and a nominally undoped
reference QW labeled with "R" (see Table~\ref{tab1} and schema in
Figs.~\ref{fig_I}, \ref{fig_II}). The samples in each set were
grown under identical conditions, excepting  the modulation doping
with Iodine donors in the barrier layer, which provides 2DEG in
the QW.

Samples \#1R and \#1D from the first set have a 67-\AA-thick ZnSe
single quantum well embedded between 1000-\AA-thick
Zn$_{0.82}$Be$_{0.08}$Mg$_{0.10}$Se barriers. To prevent the loss
of carriers escaping into the substrate and recombining at the
surface, this structure was clad between 500-\AA-thick
Zn$_{0.71}$Be$_{0.11}$Mg$_{0.18}$Se barriers with wider band gap.
Sample \#1D has two 20-\AA-thick Iodine-doped layers,
symmetrically placed in the Zn$_{0.82}$Be$_{0.08}$Mg$_{0.10}$Se
barriers and separated from the QW by a 100-\AA-thick spacers. The
electron density in this QW of $5 \times 10 ^{11}$~cm$^{-2}$ was
evaluated by the nonmonotonic behavior of the optical spectra in
the vicinity of integer filling factors \cite{17,18,23} (the
results will be collected in Fig.~\ref{fig_VII}). This electron
density corresponds to a Fermi energy of  $\varepsilon_F
=7.7$~meV. In ZnSe QWs $\varepsilon_F \mathrm{[meV]} = n_e \pi
\hbar^2 / m_e = 2.3 \times 10^{-12} \, n_e \mathrm{[cm^{-2}]} /
(m_e / m_0)$. The electron density $3 \times 10 ^{10}$~cm$^{-2}$
of in the reference sample \#1R is due to unintentional doping of
the thick barrier layers, which provides free electrons for the
single QW. The density was evaluated from the ratio of the trion
and exciton oscillator strengths \cite{22}. Note, due to possible
technological inaccuracy the well width in the set \#1R and \#1D
could be slightly different from nominal.

Samples \#2R and \#2D have a 100-\AA-thick ZnSe single QW embedded
between 1000-\AA-thick Zn$_{0.94}$Be$_{0.06}$Se barriers that are
further sandwiched between additional 500-\AA-thick
Zn$_{0.92}$Be$_{0.08}$Se barriers. The residual electron density
in \#2R is $8 \times 10 ^{10}$~cm$^{-2}$. Similar to \#1D the
modulation-doped layers in \#2D consist of 20-\AA-thick layers
symmetrically located on both sides of the well at a distance of
100~\AA. This sample has a higher electron density $1.4 \times 10
^{12}$~cm$^{-2}$ and a respectively larger Fermi energy of
$\varepsilon_F =21.5$~meV. We note here that both doped samples
satisfy the condition  $E_B^T \leq \varepsilon_F \leq E_B^X$.

Photoluminescence, PLE and reflectivity spectra were measured in
external magnetic fields applied perpendicular to the QW plane
(Faraday geometry). Circularly polarized components of these
spectra were analyzed in order to resolve states with different
spin configurations and spin orientations. Experiments in magnetic
fields up to 8~T were performed in W\"{u}rzburg in a split-coil
superconducting solenoid and for temperatures from 1.6 to 70~K. A
mid-pulse magnet ($\sim 400$-ms decay) at the National High
Magnetic Field Laboratory (Los Alamos, USA) was used for
experiments up to 48~T performed at a temperature $T=1.6$~K.
Details of the pulsed magnet setup are given in
Ref.~\onlinecite{24}. Photoluminescence was excited by UV lines of
an Ar-ion laser (W\"{u}rzburg) or by a He-Cd laser with a photon
energy of 3.8~eV (Los Alamos). For the photoluminescence
excitation measurements a dye laser (stilbene-3) was used.
Excitation density was kept below 10~W/cm$^{2}$, to exclude
considerable heating of the 2DEG. In reflectivity experiments a
halogen lamp was used as a light source. Spectra were measured by
a liquid-nitrogen-cooled charge-coupled-device detector associated
with either a 1-m or 0.3-m spectrometer.


\section{\label{sec3}OPTICAL SPECTRA AT A ZERO MAGNETIC FIELD}

First we discuss properties of PL, PLE and reflectivity spectra in
the absence of external magnetic fields, which are given in
Fig.~\ref{fig_I} for the samples \#1R and \#1D.

A PL spectrum from the undoped \#1R sample measured at $T=1.6$~K
is shown in Fig.~\ref{fig_I}a by a dashed line. It consists of two
narrow lines, each with a full-width at half-maximum (FWHM) of
1.4~meV. The linewidth is due to inhomogeneous broadening, caused
by fluctuations of the QW width and alloy fluctuations of the
barrier material. The line at 2.8259~eV labeled as $\mathrm{X}$ is
related to the 1s state of the heavy-hole exciton. The trion line
($\mathrm{T}$) is shifted by 5.1~meV to lower energy from the
neutral exciton $\mathrm{X}$ line. Identification of the charged
and neutral exciton lines is based on magneto-optical spectra (for
details see Refs.~\onlinecite{20b,21}). Both exciton and trion
transitions are also observable in reflectivity spectra (solid
lines in Fig.~\ref{fig_I}a). Due to its smaller binding energy the
trion disappears rapidly with increasing temperature in emission
and reflectivity spectra. However the exciton transition is
broadened but remains observable up to room temperature. It is
well established for QWs with low electron concentrations that the
energy difference between the exciton and trion lines increases
linearly with the Fermi energy \cite{14,21}. The electron density
of $3 \times 10 ^{10}$~cm$^{-2}$ in \#1R sample corresponds to
$\varepsilon_F = 0.5$~meV and, therefore, the binding energy of an
"isolated" trion can be evaluated as $E_B^T = 5.1 - 0.5 =
4.6$~meV.

The light-hole exciton ($X_{lh}$) in the reflectivity spectrum of
sample \#1R is shifted by 18.7~meV to higher energies with respect
to the heavy-hole exciton ($\mathrm{X}$) due to strain and quantum
confinement. The resonance of the light-hole trion ($T_{lh}$) at
2.8404~eV has a rather small oscillator strength. The binding
energy of the light-hole trion of 3.7~meV is approximately one meV
smaller than that of the heavy-hole trion which is in agreement
with earlier reports \cite{20b,21}. A weak resonance at 2.8463~eV
in the reflectivity spectrum at 30~K corresponds to the 2s-state
of the heavy-hole exciton ($X_{2s}$). The energy difference
between the 1s and 2s exciton of 21.6~meV is in a good agreement
with the calculated value of 24.7~meV for the corresponding QW
width \cite{21}.

The PL spectrum of the corresponding doped sample (\#1D) is shown
in Fig.~\ref{fig_I}b. It consists of a broad band with a maximum
at 2.810~eV and a shoulder at 2.812~eV. The FWHM of 7~meV is very
close to the value of the 2DEG Fermi energy of 7.7~meV. The PL
band has a sharp decrease of intensity at its high energy side. We
will show below that the optical transitions at this spectral
position of 2.814~eV involve electrons located in the vicinity of
the Fermi level. These transitions correspond to the
half-intensity point of the rising edge in PLE spectrum.

PLE and reflectivity spectra of sample \#1D are dominated by
strong narrow lines at 2.815 eV (labeled as $\omega_1$ in
Fig.~\ref{fig_I}b), which coincides with the high energy tail of
the PL band. The width of the resonance in the reflectivity
spectrum is about 2~meV and is equal approximately to the width of
the exciton resonance in \#1R sample. It is a remarkable fact that
at a low temperature of 1.6~K the reflectivity spectrum of the
relatively highly doped structure is indistinguishable from the
spectrum of the nominally undoped QW, where a single resonance of
the neutral exciton dominates. Both the linewidth and the
oscillator strength of the resonances in sample \#1D are very
similar to that of the exciton transition in undoped structures
\cite{15a}. Temperature dependencies presented below disclose this
similarity and reveal the difference in the origin of these
resonances. At energies higher than  the $\omega_1$ line, three
weak features can be distinguished in the PLE and reflectivity
spectra. The first peak at 2.8273~eV ($\omega_2$) is blue-shifted
by 11.8~meV with respect to the lowest absorption threshold
$\omega_1$. Note, this value is close to the  $E_B^T +
\varepsilon_F = 12.3$~meV. Transitions at 2.8332 and 2.8433~eV are
ascribed to the corresponding light-hole resonances,
${\omega_1}^{lh}$ and ${\omega_2}^{lh}$. The difference in their
transition energies is 10.1~meV. This value equals again
approximately the sum of the trion binding energy and the Fermi
energy, which gives a value of 11.4~meV for the light-hole trion.

It should be noted here that the commonly accepted notation of the
spectral features in the optical spectra of doped QWs is not
settled yet. In this paper we will follow the approach using
$\omega_1$ and $\omega_2$ resonances introduced by Cox \textit{et
al.} for the optical spectra of CdTe-based QWs in the regime
$E_B^T \leq \varepsilon_F \leq E_B^X$ \cite{18a,14}. This
notification originates from the theoretical works of Hawrylak
\textit{et al.}, where it was however introduced for the regime of
a very-high dense 2DEG with $ \varepsilon_F > E_B^X$ \cite{5,6}.
It is still an open question how far this approach and the
respective theoretical predictions can be extended to describe
optical spectra in QWs with a 2DEG of lower density. As we will
show here some of the qualitative predictions of the Hawrylak's
theory are valid to describe our experimental observations.
Namely, the theory predicts that the energy difference between two
thresholds in absorption, $\omega_1$ and $\omega_2$, corresponding
to bound states, satisfies the equality $\hbar \omega_2 - \hbar
\omega_1 = E_B^T + \varepsilon_F$ \cite{5}. This relation holds
for sample \#1D.

Optical spectra from the second set of samples (\#2R and \#2D) are
shown in Fig.~\ref{fig_II}. In most regards, they are similar to
the first set described above. The main difference comes from the
higher electron density of sample \#2D. The PL spectrum from the
reference sample \#2R consists of an exciton line ($\mathrm{X}$)
at 2.8178~eV and a trion line at 2.8131~meV with linewidths of
1.3~meV (see Fig.~\ref{fig_II}a, dashed line). Both transitions
are observable as strong resonances in PLE and reflectivity
spectra. Due to larger QW width and smaller barrier height
compared to sample \#1R, the energy splitting between the
heavy-hole and light-hole excitons is only 13.3~meV.

The shape of the PL spectrum of the doped sample \#2D is similar
to that of \#1D. The FWHM of the emission band in \#2D is
17.8~meV, which is again in a good qualitative agreement with
$\varepsilon_F = 21.5$~meV. In contrast to \#1D, the singularity
at the fundamental absorption edge $\omega_1$ of \#2D is smeared
out in the PLE spectrum and the half intensity point of the rising
edge in the absorption (PLE) is blue shifted by 3.6~meV with
respect to the high energy side of the PL band. Additionally, a
weak absorption attributed to the ${\omega_1}^{lh}$ is observable
in the PLE spectra 13~meV higher than the $\omega_1$ line energy.
In reflectivity spectra, the oscillator strength of both
transitions is weak. We were not able to distinguish a second
absorption threshold $\omega_2$ for sample \#2D. It is worthwhile
to note that even for such a high electron density  $n_e = 1.4
\times 10 ^{12}$~cm$^{-2}$, the energy position of the
characteristic absorption and emission edges in the sample \#2D
stay in the vicinity of the exciton and trion transitions of
undoped and weakly-doped samples of the same well width.

\begin{figure}[tbp]
\includegraphics[width=.31\textwidth]{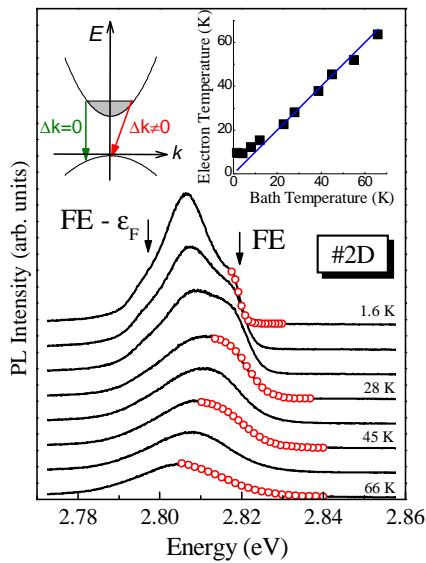}
\caption{(Color online) PL spectra of \#2D sample with $n_e = 1.4
\times 10 ^{12}$~cm$^{-2}$ at different temperatures (solid
lines). Circles represent a fitting of the high energy side of the
spectra by a Fermi distribution using the electron temperature as
a parameter. The inset shows the results of the fitting
\textit{vs} sample bath temperature. Arrows indicate indirect
transition from Fermi level ($\mathrm{FE}$), as shown in the
scheme, and from the bottom of conduction band $\mathrm{FE}
-\varepsilon_F$. } \label{fig_III}
\end{figure}

The shape of the PL spectrum of the doped samples is typical for
modulation-doped QWs with high electron densities, where the
optical transitions are interpreted as band-to-band transitions of
free carriers (see schema in Fig.~\ref{fig_III}). In this case,
all electrons in a 2DEG can recombine with photoexcited holes. It
is commonly suggested that the momentum conservation selection
rule for optical transition ($\Delta k = 0$) is partially relaxed
due to scattering with the electrons of the 2DEG, or to collective
excitations of the Fermi sea \cite{1}. As a result, indirect
optical transitions with $\Delta k \neq 0$  between the electrons
of the conduction band and holes of the valence band contribute
significantly to emission spectra. We will show below that despite
the visual similarity in the shape of emission spectra, this
simple interpretation is insufficient to describe the optical
spectra in the intermediate concentration regime where  $E_B^T
\leq \varepsilon_F \leq E_B^X$. Coulomb  interactions and
excitonic effects are of great importance in this regime.

The temperature dependence shown in Fig.~\ref{fig_III} helps to
clarify the origin of the PL band in the doped samples. It is seen
that in sample \#2D the line shape varies strongly as the bath
temperature is increased from 1.6 up to 66~K. With the temperature
increase, the  maximum becomes less pronounced and also a shoulder
on the high energy side smears out and becomes indistinguishable
for temperatures higher than 30~K. Let us first concentrate of the
shape of the high-energy tail of the emission band. We have fit it
with the Fermi function, and the results of the fit are indicated
by circles. The electron temperature, used as  a fitting
parameter, is plotted as a function of the bath lattice
temperature in the inset of Fig.~\ref{fig_III}. A solid line in
the inset is a linear dependence with unity slope, i.e. it
corresponds to the conditions when the electron temperature equals
to the bath temperature. One can see that most of the experimental
points closely follow this dependence. This allows us to conclude
that indeed the high-energy tail of the emission band in doped
2DEGs is due to  electrons in the vicinity of the Fermi level. The
Fermi edge energy derived from the fit procedure is shown in Fig.
3 by an arrow labeled $\mathrm{FE}$ at 2.819~eV for $T=1.6$~K. It
corresponds to the energy where the luminescence intensity at the
sharp high-energy side has dropped down to 50\% of its maximum
value.

\begin{figure}[tbp]
\includegraphics[width=.31\textwidth]{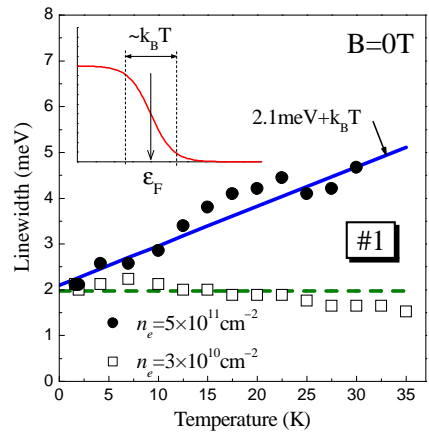}
\caption{(Color online) Temperature broadening of resonances in
reflectivity spectra for an undoped and doped ZnSe-based QWs.
Squares represent data of the exciton broadening in a sample \#1R.
Circles represent data for the $\omega_1$ resonance in \#1D. Lines
are guide to the eye. The energy distribution of electrons in
vicinity of the Fermi energy is shown schematically in the inset.
} \label{fig_IV}
\end{figure}

The linewidth of the emission band is also expected to be related
to the properties of the 2DEG, namely to the Fermi energy
$\varepsilon_F$. As can be seen from the schema in
Fig.~\ref{fig_III}, depending on the hole distribution the
emission linewidth should vary from $\varepsilon_F = 21.5$~meV up
to $(1 + m_e/m_h) \varepsilon_F = 28.5$~meV. The latter value was
calculated with electron and hole masses $m_e = 0.15 m_0$ and $m_h
= 0.46 m_0$, which have been inferred from the diamagnetic shift
of the exciton line in sample \#2R (see Ref.~\onlinecite{21}).
Experimental data agreed well with this range. For low
temperatures, where the line shape is complex, we label by an
arrow in the figure the energy position of the expected low energy
tail of the emission band at $\mathrm{FE} - \varepsilon_F$. In the
temperature range from 30 to 66~K the FWHM is 23.5~meV. The
temperature evolution of the PL band in sample \#1D is
qualitatively similar to sample \#2D and therefore the
experimental data are not shown. Modeling of the line shape is a
complicated task and is beyond the scope of this paper.

Temperature dependences are very helpful in highlighting the
qualitative differences between the resonances in reflectivity
spectra of doped and undoped samples. We note that at $T=1.6$~K,
the broadening and the amplitude (i.e. resonance oscillator
strength) of the resonances were very similar in these samples
(see Fig.~\ref{fig_I}). However, a difference in the temperature
broadening of the resonances in samples \#1R and \#1D is clearly
seen in Fig.~\ref{fig_IV}. The exciton resonance in the undoped
sample \#1R shows no temperature broadening for $T<35$~K.
Significant temperature broadening of the exciton line is expected
due to exciton scattering on LO-phonons, which becomes important
only for temperatures exceeding 100~K. In contrast, the width of
the reflectivity resonance of the doped sample \#1D increases
linearly with temperature. The slope of this increase equals $k_B
T$, which is expected for the Fermi statistics of the 2DEG in the
vicinity of the Fermi edge. This allows us to conclude that the
strong resonance  $\omega_1$ in reflectivity and PLE spectra is
due to the Coulomb interaction of the photoexcited holes and
electrons in the vicinity of the Fermi edge. In other words it is
due to an excitonic effect involving electrons at the Fermi level.

\begin{figure}[tbp]
\includegraphics[width=.47\textwidth]{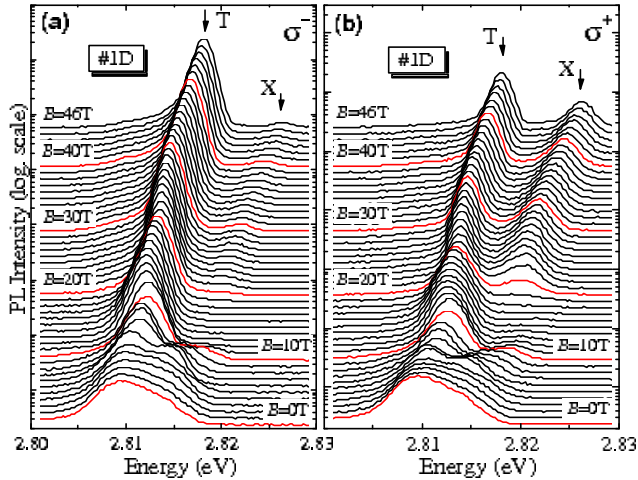}
\caption{(Color online) PL spectra from a doped 67-\AA-thick
ZnSe/Zn$_{0.82}$Be$_{0.08}$Mg$_{0.10}$Se QW \#1D with $n_e = 5
\times 10 ^{11}$~cm$^{-2}$ for magnetic fields varied from 0 to
46~T and $T=1.6$~K. (a) in $\sigma^{-}$-polarization. (b) in
$\sigma^{+}$-polarization. } \label{fig_V}
\end{figure}

Let us summarize the results of optical spectra examined in the
absence of external magnetic fields. Emission, reflectivity and
PLE spectra of the doped samples \#1D and \#2D are clearly
controlled by the 2DEG and especially by the electrons in vicinity
of the Fermi edge. Their appearance agrees qualitatively with what
is expected and commonly accepted for the highly-doped QWs with
$\varepsilon_F > E_B^X$, a situation which is often described in
terms of band-to-band optical transitions with negligible
electron-hole Coulomb interactions. However, we study here II-VI
heterostructures with strong Colomb  interactions and also examine
an intermediate electron density regime characterized by $E_B^T
\leq \varepsilon_F \leq E_B^X$. Is it clear from our experimental
data that the high energy side of the emission spectra and the
absorption edge, detected via reflectivity and PLE spectra, does
not correspond to the energy of the optical transition from the
valence band to the Fermi edge. These characteristic energies are
moved to lower energies by about 30~meV, which is the exciton
binding energy, suggesting the importance of electron-hole Coulomb
interactions. Schematically these characteristic energies are
displayed in Fig.~\ref{fig_XII}. We will discuss this schema and
designation of the characteristic energies in more detail after
presenting the results of magnetic field studies.


\section{\label{sec4}MODIFICATION OF OPTICAL SPECTRA IN HIGH MAGNETIC FIELDS}

\subsection{\label{sub4A} PL spectra}

\begin{figure}[tbp]
\includegraphics[width=.43\textwidth]{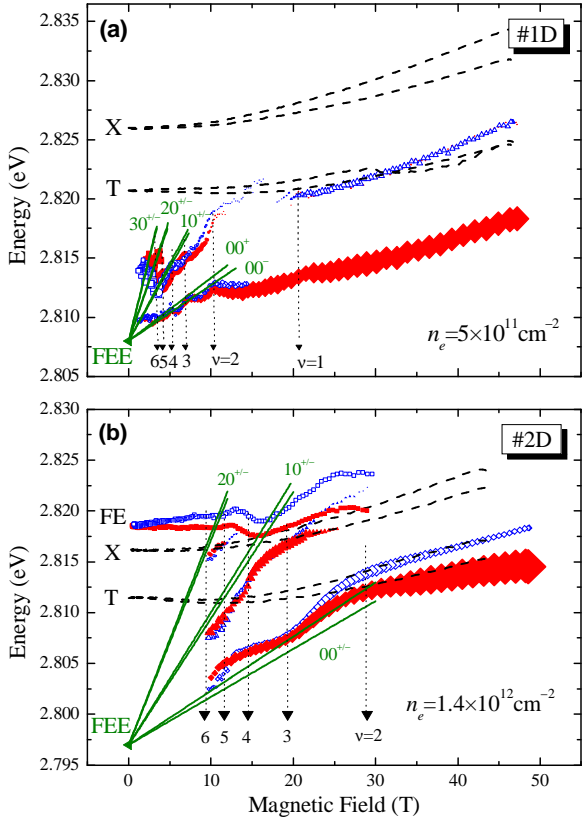}
\caption{(Color online) (a) Energy of PL maxima of a doped
67-\AA-thick ZnSe/Zn$_{0.82}$Be$_{0.08}$Mg$_{0.10}$Se QW \#1D with
$n_e = 5 \times 10 ^{11}$~cm$^{-2}$ \textit{vs} magnetic field
detected in $\sigma^{+}$ (open symbols) and $\sigma^{-}$ (closed
symbols) polarizations. Exciton ($\mathrm{X}$) and trion
($\mathrm{T}$) diamagnetic shifts for \#1R sample are shown by
lines. Landau level fan chart is shown for off-diagonal
transitions with in-plane electron and hole effective masses of
$m_e = 0.15 m_0$ and of $m_h = 0.46 m_0$, respectively. Vertical
lines indicate magnetic fields of integer filling factors. The
fundamental emission edge ($\mathrm{FEE}$) at 2.808~eV corresponds
to the low energy side of the luminescence band at $B=0$. (b) Same
for a doped 100-\AA-thick ZnSe/Zn$_{0.94}$Be$_{0.06}$Se QW \#2D
with $n_e = 1.4 \times 10 ^{12}$~cm$^{-2}$ and the corresponding
reference sample \#2R. Squares represent the energy of the
transitions of electrons at the Fermi level ($\mathrm{FE}$). The
fundamental emission edge ($\mathrm{FEE}$) is located at
$E_{\mathrm{FEE}} = 2.797$~eV. It is equal to the low energy side
of the luminescence band $\mathrm{FE} - \varepsilon_F$ at $B=0$
(see also Fig.~\ref{fig_III}). $T=1.6$~K.  } \label{fig_IX}
\end{figure}

We turn now to modifications of optical spectra in external
magnetic fields. The evolution of the photoluminescence spectra
from sample \#1D is shown in Fig.~\ref{fig_V} for  $\sigma^{-}$
(a) and $\sigma^{+}$-polarization (b). PL intensity is given in a
logarithmic scale. The broad PL band that is characteristic at low
magnetic fields (see also Fig.~\ref{fig_I}b) is transformed for
$B>20$~T into two narrow lines which are very similar to the
exciton and trion lines of the reference QWs. The polarization of
the higher-energy line, which resembles the exciton in the undoped
structures, is opposite to the polarization of the trion. At low
magnetic fields, the PL band splits into a set of lines, which
shift almost linearly to higher energies with increasing field.
Above 10.3~T, the linear energy shift of the lowest line converts
into a diamagnetic shift typical for excitonic states. Details of
this transformation are easier to follow on the fan-chart diagram
given in Fig.~\ref{fig_IX}a.  Similar behavior has been reported
recently for GaAs-based QWs \cite{11,12}. It was shown that the
transformation occurs at a filling factor $\nu=2$ and that at high
magnetic fields the emission is indistinguishable from trions.
Theoretically, the transformation from a linear shift for $\nu>2$
to a diamagnetic shift for $\nu<2$ can be explained by a hidden
symmetry, which can hold for $\nu<2$ \cite{8}, and the energy of
optical transitions for $\nu<2$ resembles the energy of the
exciton.

\begin{figure}[tbp]
\includegraphics[width=.47\textwidth]{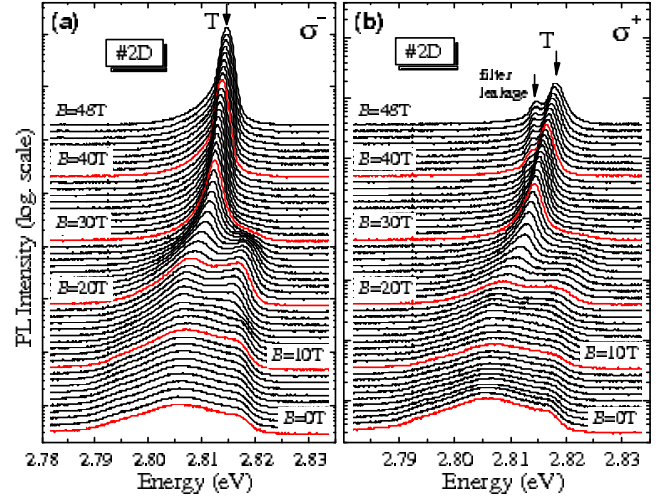}
\caption{(Color online) PL spectra from a doped 100-\AA-thick
ZnSe/Zn$_{0.94}$Be$_{0.06}$Se QW \#2D with $n_e = 1.4 \times 10
^{12}$~cm$^{-2}$ for magnetic fields varied from 0 to 46~T and
$T=1.6$~K. (a) in $\sigma^{-}$-polarization. (b) in
$\sigma^{+}$-polarization. } \label{fig_VI}
\end{figure}

Figures~\ref{fig_VI} and \ref{fig_IX}b show the influence of the
magnetic field on the luminescence spectra of the sample \#2D.
Field induced changes of the spectra are similar to that of sample
\#1D. As a consequence of higher electron concentration, the
characteristic transformation of the spectra appears at higher
magnetic fields. For low magnetic fields, the high energy side of
the luminescence band shows oscillations, due to the depopulation
of higher Landau levels. These oscillations became very pronounced
for filling factor  $\nu=4$ at $B=14.5$~T. For filling factor
$\nu=2$ at 29~T the trion-like luminescence is recovered. We do
not reach here the regime of filling factor $\nu=1$, which is
expected at $B=58$~T, i.e. beyond the maximum field available. As
a consequence we do not observe the recovery of the exciton-like
emission line, as was observed for sample \#1D with lower electron
density.

\begin{figure}[tbp]
\includegraphics[width=.42\textwidth]{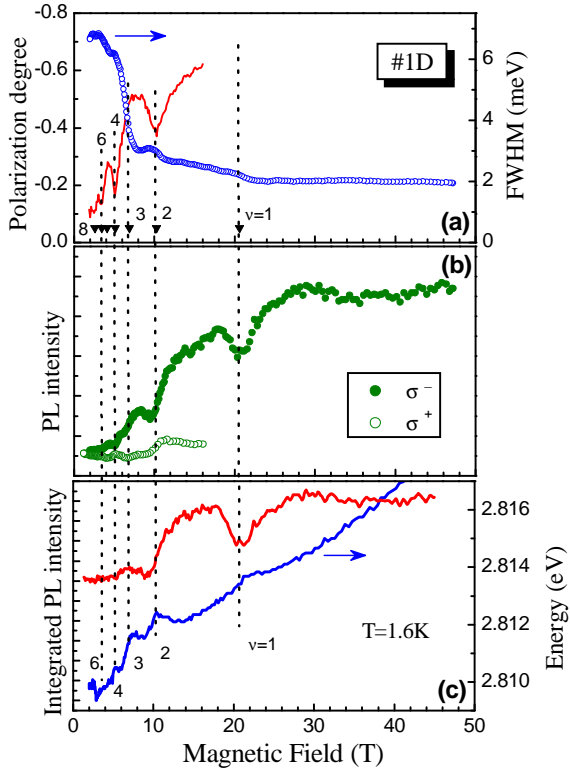}
\caption{(Color online) Summary of nonmonotonic behavior of the
magneto-optical properties of a doped 67-\AA-thick
ZnSe/Zn$_{0.82}$Be$_{0.08}$Mg$_{0.10}$Se QW \#1D with $n_e = 5
\times 10 ^{11}$~cm$^{-2}$: (a) Polarization degree and linewidth
of PL band; (b) PL intensity for two circular polarizations; (c)
Integrated emission intensity over both polarizations and energy
shift of PL maxima detected in $\sigma^{-}$ polarization. Dotted
lines indicate locations of integer filling factors.  }
\label{fig_VII}
\end{figure}

The 2DEG density in the modulation-doped samples was evaluated
from oscillations of the optical properties at integer filling
factors. With growing magnetic field the number of electrons in
the uppermost occupied Landau level varies periodically. As a
result many-body effects such as the electron exchange and
correlation energy, the screening of the Coulomb interactions of
the electron and hole, and the degree of spin polarization of the
2DEG oscillate with increasing magnetic field. The oscillations
can be found in a variety of optical properties, such as the
degree of circular polarization, the integrated intensity of
luminescence, and the energy positions of the lines. These
oscillations are especially pronounced in the vicinity of integer
filling factors of a two-dimensional electron gas. In the
structures studied here, critical behavior at integer filling
factors has been established for different characteristics of
luminescence. For sample \#1D the oscillations are collected in
three panels of Fig.~\ref{fig_VII}. The energy shift of the lowest
PL maximum shows an upward cusp at integer filling factors from 1
to 4, and downward convex curves between them. The most pronounced
feature at $B=10.3$~T corresponds to $\nu=2$. At this magnetic
field the linear shift of the PL maxima changes to the diamagnetic
behavior typical for trions. From the magnetic field value
corresponding to $\nu=2$ we derive the 2DEG density $n_e = 5
\times 10 ^{11}$~cm$^{-2}$ and calculate the expected fields for
the set of integer filling factors. These fields are marked by
vertical dotted lines in panels (a), (b) and (c). They are in good
agreement with evaluations of $\varepsilon_F$ and $n_e$ derived
from the linewidth of emission line, which however are much less
accurate. This also coincides well with data on the optical
detection resonance spectroscopy with the use of far-infrared
radiation performed for the same sample \cite{18}.

\begin{figure}[tbp]
\includegraphics[width=.43\textwidth]{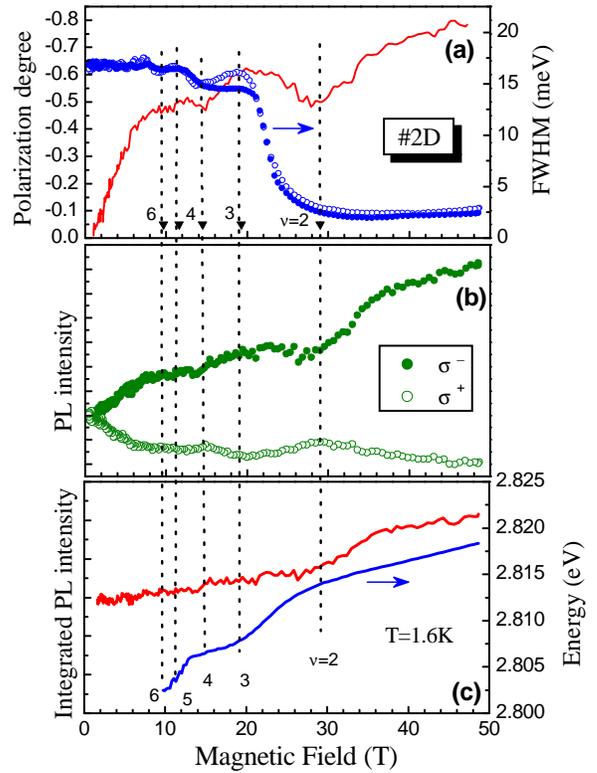}
\caption{(Color online) Summary of nonmonotonic behavior of the
magneto-optical properties of a doped 100-\AA-thick
ZnSe/Zn$_{0.94}$Be$_{0.06}$Se QW \#2D with $n_e = 1.4 \times 10
^{12}$~cm$^{-2}$: (a) Polarization degree and linewidth of
luminescence band; (b) PL intensity for two circular polarization;
(c) Integrated emission intensity over both polarizations, and
energy shift of PL maxima detected in polarization. Dotted lines
indicate locations of integer filling factors. } \label{fig_VIII}
\end{figure}

Very pronounced features are observed for the even filling factors
$\nu = 2,4,6$ for the PL circular polarization degree, which
exhibits a minimum when the number of Landau levels that are fully
occupied for each spin are equal (Fig.~\ref{fig_VII}a). Also, the
PL intensity in $\sigma^{+}$-polarization shows dips at  $\nu=1$,
2 and 4 (Fig.~\ref{fig_VII}b) and maxima at $\nu=2$ and 4 in
$\sigma^{+}$-polarization. The integrated PL intensity also shows
oscillatory behavior (Fig.~\ref{fig_VII}c), which can be explained
by a decrease of the radiative recombination rate at integer
filling factors. However, this interpretation requires additional
experimental support, e.g. time-resolved measurements of the
emission decay. It is worthwhile to note here that oscillatory
behavior was also detected in experiments with nonresonant heating
of a 2DEG by a far-infrared laser (184.3~$\mu$m) \cite{18}. Clear
minima in the electron temperature were found for  $\nu=2$ and 4,
i.e. when electron acceleration within the same Landau level is
prohibited because all the states are occupied.
Figure~\ref{fig_VIII} shows a summary of the oscillating
properties for the sample \#2D. In contrast to sample \#1D,
oscillations of the lowest transition energy and of the total PL
intensity are less pronounced. Landau level broadening in the
sample \#2D is considerable, which indicates strong scattering of
electrons by impurities and by electrostatic potential of ionized
donors from modulation doped layers. Luminescence of the different
Landau levels can be clearly separated only for magnetic fields
above 15~T. Nevertheless, the 2DEG density can be derived by the
analysis of luminescence linewidth, which oscillates in phase with
the Fermi energy for each electron spin in the corresponding
circular polarization (see Fig.~\ref{fig_VIII}a,b). The degree of
circular polarization given in Fig.~\ref{fig_VIII}a also
oscillates and shows pronounced minima at even filling factors and
maxima at odd filling factors.

We now relate the experimental data from Fig.~\ref{fig_IX} to a
simple fan chart diagram for transitions between Landau levels of
conduction and valence bands. The energy positions of these
transitions, labeled as ${N_e N_h}^{\, \sigma}$ are expected at
\begin{equation}
E_{N_e N_h}^{\sigma} = E_{\mathrm{FEE}} + (N_e + \frac{1}{2})\hbar
\omega_{ce}  + (N_h + \frac{1}{2})\hbar \omega_{ch} + E_z^\sigma
\,. \label{eq1}
\end{equation}
Here $E_{\mathrm{FEE}}$ is the fundamental emission edge, which
includes the band gap renormalization and the excitonic effect
between electrons and photoholes. $E_z^\sigma$ accounts for the
Zeeman splitting of the valence and conduction band states. We use
here the respective values determined for the reference samples
and do not account for the possible modification of the electron
Zeeman splitting due to electron-electron exchange interaction
\cite{16}. $\hbar \omega_{ce} = \hbar e B / m_e$ and $\hbar
\omega_{ch} = \hbar e B / m_h$ are the cyclotron energies of the
conduction band electrons and the valence band holes,
respectively, and $N_{e(h)} = 0,1,2...$ is the Landau level
number. While the selection rule $N_e = N_h$ pertains for light
absorption (i.e. for PLE and reflectivity spectra), it can be
violated for emission spectra by carrier scattering with the Fermi
sea electrons.

In Fig.~\ref{fig_IX} fan diagrams of emission lines in external
magnetic fields are shown for two doped samples. The size of the
symbols corresponds to the emission intensity. The symbol size at
higher Landau levels ($N_e > 0$) is 5 times magnified with respect
to $N_e = 0$. In low magnetic fields for  $\nu>2$ the experimental
data of both samples are described considerably well by
Eq.~(\ref{eq1}) by supposing optical transitions between electrons
from Landau levels $N_e =0, 1, 2, 3...$ and holes from Landau
level $N_h=0$. The effective masses $m_e = 0.15 m_0$ and $m_h =
0.46 m_0$ were used for the calculation of emission energy given
by solid lines, and $E_{\mathrm{FEE}}$ was treated as a fitting
parameter. The best agreement with the experimental data for the
sample \#1D is achieved for $E_{\mathrm{FEE}}=2.808$~eV. This
value coincides reasonably well with the low energy tail of the
emission band at a zero field determined as $\mathrm{FE} -
\varepsilon_F = 2.8065$~eV (see also Fig.~\ref{fig_I}).

For the \#1D sample the transformation to diamagnetic behavior at
$\nu < 2$ takes place at $B = 10.3$~T. It is instructive to
compare in this regime emission spectra of the doped and reference
samples. Modification of the magneto-optical spectra in ZnSe-based
QWs with a low-density 2DEG has been studied in details in Refs.
\onlinecite{21,25}. Dashed curves in Fig.~\ref{fig_IX}a represent
the shift of exciton and singlet trion states, measured for the
sample \#1R. The diamagnetic shift of these lines (i.e. quadratic
with increasing magnetic fields) is characteristic for excitons
and trion complexes. It is worthwhile to note that the properties
of the reference structure (energy shift, PL intensity,
polarization degree) change smoothly with magnetic field, showing
none of the cusps or jumps that are typical for doped samples. One
can see that the strongest line of the sample \#1D shifts parallel
to the trion line of the reference sample in wide field range from
10.3 up to 48~T corresponding to $\nu < 2$. Also the second line,
which appears for the sample \#1D at  $\nu < 1$ shows a
diamagnetic shift that is parallel with the excitonic line of the
reference sample. The energy offset between transitions in the
doped and reference samples are 8~meV. As we will see below it is
negligible for the second set of samples \#2R and \#2D, therefore
we believe that for the offset between samples \#1R and \#1D is
mainly caused by the difference in QW widths caused by a growth
inaccuracy.

Figure~\ref{fig_IX}b shows the corresponding data for sample \#2D
with an electron density of $n_e = 1.4 \times 10 ^{12}$~cm$^{-2}$.
The situation is similar to the case of sample \#1D. In low
magnetic fields for $\nu > 2$ transitions energies show a
reasonable agreement with the fan chart diagram. Due to the large
broadening of the Landau levels the energy position of the
transitions between the Landau levels cannot be resolved for
higher filling factors $\nu > 6$. Energies of the high-energy
cutoff of emission spectra (that involves electrons in vicinity of
the Fermi edge), are shown by squares. It is remarkable that again
the fundamental emission edge $E_{\mathrm{FEE}}=2.797$~eV is equal
to the low energy shoulder of the PL band $\mathrm{FE} -
\varepsilon_F = 2.7975$~eV (see also Figs.~\ref{fig_II} and
\ref{fig_III}). Such a systematic behavior from both doped samples
confirms our phenomenological description given in
Fig.~\ref{fig_XII}.

In magnetic fields above 28~T ($\nu < 2$) the energy of main
optical transition of \#2D coincides with the trion energy in \#2R
(compare symbols with dashed lines in Fig.~\ref{fig_IX}b). The
recovery of the exciton transition cannot be found in this plot as
$\nu = 1$ is expected at $B=58$~T which exceeds the highest field
available for the used experimental facility.

\subsection{\label{sub4B} Reflectivity spectra}

\begin{figure}[tbp]
\includegraphics[width=.47\textwidth]{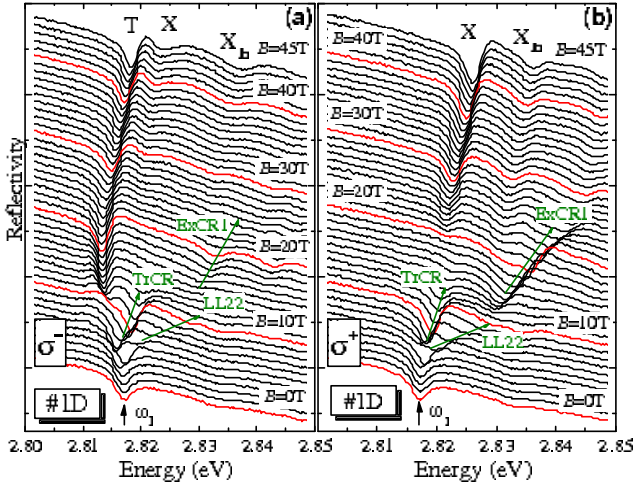}
\caption{(Color online) Reflectivity spectra of a doped
67-\AA-thick ZnSe/Zn$_{0.82}$Be$_{0.08}$Mg$_{0.10}$Se QW \#1D with
$n_e = 5 \times 10 ^{11}$~cm$^{-2}$ in a magnetic field range from
0 to 45~T at $T=1.6$~K. (a) $\sigma^{-}$-polarization. (b)
$\sigma^{+}$-polarization. } \label{fig_X}
\end{figure}

We turn now to the discussion of the reflectivity spectra of the
doped samples in external magnetic fields. The spectra from sample
\#1D in magnetic fields are given in Fig.~\ref{fig_X}. The shorter
data acquisition times required by the pulsed field experiments do
not allow resolution of the weak  $\omega_2$ resonances visible in
Fig.~\ref{fig_I}b. At $B=0$, the reflectivity spectra are
dominated by the $\omega_1$ resonances of the heavy-hole and
light-hole transitions at 2.8169 and 2.8331 ~eV, respectively. In
the high field limit, two heavy-hole transitions can be clearly
seen, analogous to the exciton and trion resonances in the
reference structure \#1R. Additionally, the light-hole transition
gains oscillator strength for both polarizations. In the
following, we focus our discussion on the heavy-hole transitions.
In $\sigma^{-}$-polarization, the reflectivity spectra at high
fields are dominated by the trion-like resonance. A weaker
exciton-like resonance is recovered at filling factors $\nu < 1$.
In $\sigma^{+}$-polarization only the exciton-like resonance is
restored for filling factors  $\nu < 1$. Polarization properties
of the exciton-like and trion-like resonances coincide with the
behavior of the reference sample. The strong circular polarization
of the transitions originates from the complete spin polarization
of the 2DEG for filling factor $\nu < 1$. In this regime all
electrons occupy states with spin $-1/2$. Trion singlet states
have antiparallel electron configuration. As a result, this state
can only be excited resonantly with  $\sigma^{-}$-polarized light
(for details see Ref.~\onlinecite{20b}).

\begin{figure}[tbp]
\includegraphics[width=.47\textwidth]{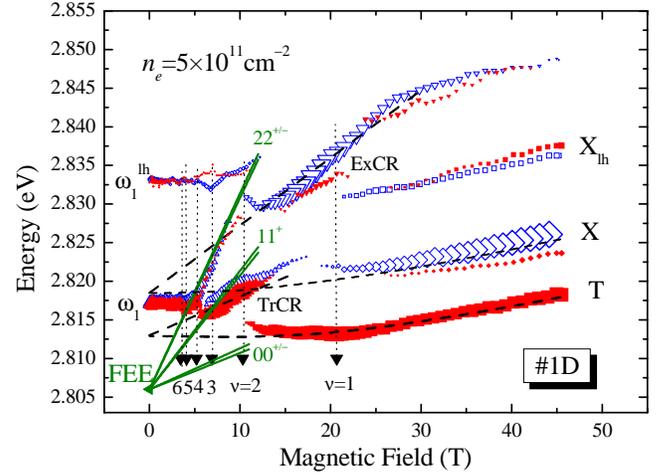}
\caption{(Color online) Energies of reflectivity lines \textit{vs}
magnetic field detected in  $\sigma^{+}$ (open symbols) and
$\sigma^{-}$ (closed symbols) polarizations for a doped
67-\AA-thick ZnSe/Zn$_{0.82}$Be$_{0.08}$Mg$_{0.10}$Se QW \#1D with
$n_e = 5 \times 10 ^{11}$~cm$^{-2}$ . Intensities of resonances
are displayed by the size of the symbols. Solid lines represent
Landau level fan chart for diagonal transitions with in-plane
electron and hole effective masses of $m_e = 0.15 m_0$ and of $m_h
= 0.46 m_0$, respectively. Vertical lines indicate locations of
integer filling factors. The energy of the renormalized band gap
at 2.8065~eV obtained from the extrapolation to zero field
corresponds well to the energy position  $\hbar \omega_1 - (1+m_e
/ m_h) \varepsilon_F =2.8053$~eV. Dashed curves represent the
diamagnetic shift of the trion and exciton as well as the energy
of the combined $\mathrm{TrCR}$ and $\mathrm{ExCR}$ processes. }
\label{fig_XI}
\end{figure}

At intermediate magnetic fields the $\omega_1$ line transforms
into a set of resonances, which shift approximately linearly with
magnetic field. The origin of these resonances is discussed in the
following. Figure~\ref{fig_XI} shows resonance energies of
reflectivity lines \textit{vs} magnetic fields (symbols). Size of
the symbols corresponds to the relative oscillator strength of the
resonances. At low magnetic fields, the reflectivity spectra are
dominated by the $\omega_1$ resonance. At $\nu=5$, when the
$N_e=2$ Landau level for electrons with spin $+1/2$ becomes
depopulated, the $\omega_1$ resonance starts shifting linearly to
higher energies and loses its oscillator strength in
$\sigma^{+}$-polarization. The slope of this shift is 2.5~meV/T.
This value is in good agreement with the expected energy shift of
diagonal transitions between electron and hole Landau-levels with
$N_{e(h)}=2$, which should have a slope of 2.55~meV/T for $m_e =
0.15 m_0$ and of $m_h = 0.46 m_0$. Analogous to the PL data, it is
instructive to compare the reflectivity data to a simple fan chart
diagram derived from Eq.~(\ref{eq1}). The best agreement with the
experimental data was achieved using
$E_{\mathrm{FEE}}=2.8065$~eV. This value is very close to the
expected energy position $\hbar \omega_1 - (1+m_e / m_h)
\varepsilon_F =2.8053$~eV.

The calculated transitions between Landau levels with $N_e = N_h$
are shown in Fig.~\ref{fig_XI} by solid lines. At $\nu = 4$, where
only two Landau levels for each spin are occupied, the main
resonance shows a small redshift in  $\sigma^{-}$-polarization and
$\sigma^{+}$-polarization, respectively. At $2 < \nu < 4$, the
main resonances shift approximately linearly with slopes of
0.55~meV/T and 0.72~meV/T for the respective polarizations. These
values deviate strongly from the expected slope of  $N_e = N_h =1$
transitions, which is predicted to be about 1.5~meV/T. In samples
with lower electron density at filling factor $2 < \nu < 4$ a
combined process is observable in absorption and reflectivity in
which an electron is excited from a filled Landau level to a
higher empty Landau level during the creation of a trion. This
four-particle process is known as a combined trion cyclotron
resonance ($\mathrm{TrCR}$) and has a slope of about $\frac{1}{2}
\hbar \omega_{ce}$ \cite{27,28}. The energies of the
$\mathrm{TrCR}$ line, extrapolated to a zero field, meets the
energy of the "bare" trion, i.e. the energy obtained from
extrapolation of the diamagnetic shift of the trion to low
magnetic fields (dashed line). This line should therefore have a
trion origin.

At $\nu \leq 2$  a new resonance gains oscillator strength in the
reflectivity spectra, which is blueshifted with respect to the
exciton resonance. From the slope it can be identified as a
combined exciton-cyclotron resonance ($\mathrm{ExCR}$) \cite{26}.
This resonance is due to a process in which photocreation of a
neutral exciton occurs simultaneously with a transition of a
background electron between Landau levels. The $\mathrm{ExCR}$
line shifts linearly with magnetic fields as 0.9~meV/T (dashed
line). Extrapolating to the zero-field limit, the $\mathrm{ExCR}$
line meets 2.8185~eV, which is the energy of the extrapolation of
the diamagnetic shift of the exciton, as illustrated by a dashed
line in Fig.~\ref{fig_XI}. The $\mathrm{ExCR}$ slope of 0.9~meV/T
is very close to 0.96~meV/T derived from the theoretical approach
of Ref.~\onlinecite{26}, as $(1 + m_e /(m_e + m_h)) \hbar
\omega_{ce}$ with $m_h = 0.46 m_0$. As already mentioned, the size
of the symbols in Fig.~\ref{fig_XI} reflects the relative
oscillator strength of the resonances. One can see that the
$\mathrm{ExCR}$ process gains its maximum oscillator strength
close to the filling factor $\nu = 1$, where the probability to
find one (only one) electron in the orbit of photogenerated
exciton is maximal. We should note here that in these experiments
the exciton Bohr radius is still 1.5-2.5 times smaller than the
magnetic length of electrons, which is 57~{\AA}  at 20~T.

Summarizing, reflectivity data show clear evidence of many body
interactions in low magnetic fields. The $\omega_1$ resonance at
the fundamental absorption edge transforms to a Landau-level
behavior with  $N_e= N_h$. In high magnetic fields, however,
exciton and trion resonances are restored in reflectivity spectra,
and cyclotron shifts of the observed lines are well described in
terms of combined $\mathrm{ExCR}$ and $\mathrm{TrCR}$ processes.


\section{\label{sec5}PHENOMENOLOGICAL PICTURE AND CONCLUSIONS}

There are a number of numerical factors which must be taken into
account in order to quantitatively model the  magneto-optical
spectra of QWs containing a 2DEG. Many of these factors result
from  many-body effects, and cannot, therefore, be treated
analytically. Among them are band-gap renormalization, screening
of excitonic states, and perturbation of the 2DEG by the presence
of a photohole. Performing such calculations is beyond the
intended scope of this paper. We are not aware of any such
calculations for II-VI QWs. However, one can formulate a
phenomenological description of the observed optical spectra and
their modification in magnetic fields for the regime $E_B^T \leq
\varepsilon_F \leq E_B^X$. We present such a description here with
an aim to qualitatively summarize the main trends observed in
2DEGs  with strong Coulomb interactions.

\begin{figure}[tbp]
\includegraphics[width=.45\textwidth]{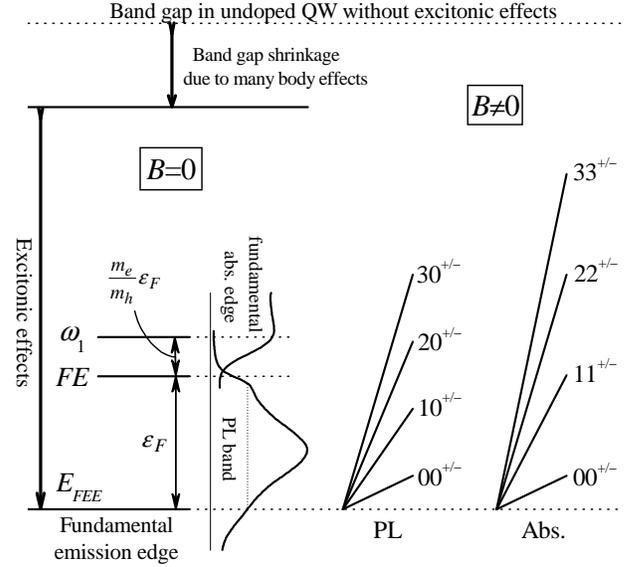}
\caption{Energetic hierarchy summarized from experimental data,
which are obtained for QWs containing a 2DEG of moderate density.
} \label{fig_XII}
\end{figure}

Figure~\ref{fig_XII} summarizes a phenomenological picture of
optical transitions and their evolution in magnetic field. The
starting point in the scheme is the band gap of the undoped QW,
i.e. the QW without a 2DEG and without any photocarriers. The
presence of the 2DEG leads to band gap renormalization, i.e. to a
decrease of the band gap which results from repulsive
electron-electron interactions. In the presence of photoholes,
excitons are formed, which are not fully screened in the
considered regime of low and moderate electron densities. The
lowest possible transition in emission will involve an electron
from the very bottom of the 2DEG. It is denoted in the diagram as
the fundamental emission edge ($\mathrm{FEE}$). It is convenient
to relate all other characteristic energies and spectral shifts to
the $E_{\mathrm{FEE}}$, as shown. The observed emission band
arises from all electrons in the 2DEG, since the optical selection
rules are partially relaxed and indirect transitions with $\Delta
k \neq 0$ become  possible. The upper edge of the emission band
results from electrons at the Fermi edge (this optical transition
is labeled as "$\mathrm{FE}$" in Fig.~\ref{fig_XII}), and it is
shifted by $\varepsilon_F$ to higher energy from the
$\mathrm{FEE}$. For the case of optical absorption, direct
transitions ($\Delta k = 0$) possessing stronger oscillator
strength dominate. An absorbed photon moves an electron from the
valence band into the empty states in conduction band which are
just above the Fermi energy of the 2DEG. Due to the finite hole
mass, this absorption transition associated with the Fermi edge is
blueshifted by $(1 + m_e / m_h) \varepsilon_F$ from the
$\mathrm{FEE}$, or correspondingly, by $(m_e / m_h) \varepsilon_F$
from the emission transition labeled "$\mathrm{FE}$" in
Fig.~\ref{fig_XII}. In the reflectivity and PLE spectra shown in
this paper, this absorption transition leads to a fundamental
absorption edge and is denoted by $\omega_1$.

Magneto-optical spectra can also be qualitatively explained in the
same terms (Fig.~\ref{fig_XII}). Magnetic field splits the
conduction and valence bands into Landau levels. Emission spectra
arise from transitions between occupied Landau levels in the
conduction band ($N_e = 0, 1, 2,...$) to the zeroth Landau level
in the valence band ($N_h = 0$). They appear as local maxima in
the emission band, which shift linearly with magnetic field.
Extrapolating their shifts to zero magnetic field shows
convergence on the $\mathrm{FEE}$ energy. In absorption spectra
(i.e., reflectivity and PLE), only transitions with $N_e = N_h$
should become visible above  $\omega_1$, which implies transitions
from the valence band to the unoccupied states in the conduction
band. Extrapolating their energy shifts to zero field again
exhibits convergence upon the $\mathrm{FEE}$ energy, similar to
the emission lines. This phenomenological picture is valid for
lower magnetic fields such that more than one Landau level is
occupied by electrons, i.e. for filling factor  $\nu > 2$.

In high magnetic fields, where  $\nu < 2$, optical spectra are
very similar to those from  QWs with a very low electron density.
One can see trion-like and exciton-like transitions shifting
diamagnetically with increasing field. Also, combined processes of
exciton-cyclotron and trion-cyclotron resonances become visible.

In conclusion, we have presented a detailed experimental study of
the PL, PLE and reflectivity spectra of ZnSe-based QWs containing
2DEGs at various temperatures and in high magnetic fields. We
focused here on modulation-doped QWs with moderate 2DEG density,
such that the Fermi energy falls in the range between the trion
and exciton binding energies $E_B^T \leq \varepsilon_F \leq
E_B^X$. To our knowledge, a quantitative theory that describes the
magneto-optical spectra of such 2DEGs does not yet exist.
According to the experimental data presented in this paper, such a
theory must account for excitonic effects that are modified by the
many body response of the 2DEG. We hope that our findings will
encourage theoretical efforts toward a better understanding of
this challenging field.


\begin{acknowledgments}

The authors are thankful to H.~A.~Nickel and B.~D.~McCombe for
collaboration at initial stages of this study and to
V.~P.~Kochereshko for useful discussions. The work was supported
by the Deutsche Forschungsgemeinschaft (SFB 410).

\end{acknowledgments}


\end{document}